\documentclass[two-column, nofootinbib]{revtex4-1}

\usepackage{amsmath}
\usepackage{tabularx}
\usepackage{graphicx}
\usepackage[usenames,dvipsnames,svgnames]{xcolor}
\usepackage{hyperref}

\hypersetup{dvips,dvipdfm,colorlinks=true,urlcolor=magenta,filecolor=magenta,linktoc=page,citecolor=red,linkcolor=blue,bookmarks=true}

\begin{document}

\title{A closer look at the Barboza-Alcaniz equation of state parametrization}

%\author{}}
%\affiliation{}

\author{Anindita Mondal\footnote{Electronic Address: \texttt{\color{blue}anindita12@bose.res.in}}}
\affiliation{Department of Astrophysics \& Cosmology, \\ S. N. Bose National Centre for Basic Sciences, Kolkata 700106, West Bengal, India.}

\author{Subhajit Saha\footnote {Electronic Address: \texttt{\color{blue} subhajit1729@gmail.com}}}
\affiliation{Department of Mathematics, \\ Panihati Mahavidyalaya, \\ Sodepur 700110, West Bengal, India.}

%%%%%%%%%%%%%%%%%%%%%%%%%%%%%%%%%%%%%%%%%%%%%%%%%%%%%%%%%%%%%%%%%%%%%%%%%%%%%%%%%%%%%%%%%%%%%%%%%%%%%%%%%%%%%%%%%%%%%%%%%%%%

\begin{abstract}

The Barboza-Alcaniz EoS parametrization has been considered and its $q$-parametrization has been investigated in search for a thermodynamic motivation. For this, we have studied the validity of the generalized second law of thermodynamics as well as the thermodynamic equilibrium considering the cosmological apparent horizon as the boundary. Also, an expression for the particle creation rate has been obtained in terms of $q$ assuming an adiabatic particle creation scenario and its behavior has been studied for consistency during various phases of evolution of the Universe as suggested by various thermodynamic arguments found in the literature.

\end{abstract}

\maketitle

%%%%%%%%%%%%%%%%%%%%%%%%%%%%%%%%%%%%%%%%%%%%%%%%%%%%%%%%%%%%%%%%%%%%%%%%%%%%%%%%%%%%%%%%%%%%%%%%%%%%%%%%%%%%%%%%%%%%%%%%%%%%

%%%%%%%%%%%%%%%%%%%%%%%%%%%%%%%%%%%%%%%%%%%%%%%%%%%%%%%%%%%%%%%%%%%%%%%%%%%%%%%%%%%%%%%%%%%%%%%%%%%%%%%%%%%%%%%%%%%%%%%%%%%%

\section{Introduction} 

Parametrization of the deceleration parameter $q$ has been found to be a useful tool towards a more complete characterization of the evolutionary history of the Universe. In the literature, several well-known $q$-parametrizations have been proposed \cite{q1,q2,q3,q0,q4,q5,q6,q7,q8,q9,q10,q11,q12,q13}. In this short paper, we shall undertake a study of the Barboza-Alcaniz EoS parametrization \cite{Barboza1}. With the corresponding $q$-parametrization, we shall study the validity of the generalized second law (GSL) of thermodynamics and thermodynamic equilibrium (TE). Moreover, we shall obtain the expression for the particle creation rate in the gravitationally induced isentropic\footnote{The words "isentropic" and "adiabatic" can be used interchangeably.} particle creation scenario and investigate the consistency in its behavior based on thermodynamic arguments found in the literature.
  
We shall consider a universe governed by the flat, homogeneous and isotropic Friedmann-Lemaitre-Robertson-Walker (FLRW) metric in comoving coordinates ($t$, $r$, $\theta$, $\phi$) as
\begin{equation} \label{flrw}
ds^2=-dt^2+a^2(t)\left[dr^2+r^2(d\theta ^2+\text{sin}^2\theta d\phi ^2)\right],
\end{equation}
where $a(t)$ is the scale factor of the Universe. Assuming the energy-momentum tensor $T_{\mu \nu}$ of the form 
\begin{equation}
T_{\mu \nu}=(\rho +p)u_{\mu}u_{\nu}+pg_{\mu \nu}
\end{equation}
with $u^{\mu}$ as the 4-velocity of the fluid. The Friedmann and the acceleration equations can be derived from the Einstein's field equations as
\begin{equation} \label{fa}
H^2=\frac{8\pi G}{3}\rho~~~~~~~~\text{and}~~~~~~~~\dot{H}=-4\pi G(\rho +p)
\end{equation}
respectively, where $H=\frac{\dot{a}}{a}$ is the Hubble parameter, whereas $\rho$ and $p$ denote, respectively, the energy density and the pressure of the cosmic fluid. The energy-momentum conservation equation can be obtained using the expressions in Eq. (\ref{fa}) and it has the form
\begin{equation} \label{emce}
\dot{\rho}+3H(\rho +p)=0.
\end{equation}
The deceleration parameter $q$ is defined as
\begin{eqnarray} \label{q}
q &=& -\frac{\dot{H}}{H^2}-1 \nonumber \\
&=& \frac{3}{2}\left(1+\frac{p}{\rho}\right)-1 \nonumber \\
&=& \frac{1}{2}(1+3w).
\end{eqnarray}
$w=\frac{p}{\rho}$ is the equation of state (EoS) of the cosmic fluid chosen. It may be either a constant or a variable function of the scale factor $a$, or equivalently, the redshift parameter $z$ ($=-1+1/a$). A positive $q$ indicates deceleration while $q<0$ implies acceleration.

As stated earlier, in the present work, we shall focus our investigation on a particular form of the EoS parameter $w(z)$, popularly known as the Barboza-Alcaniz parametrization. The explicit form is given by \cite{Barboza1}
\begin{equation}
w(z)=w_0 + w_1 \frac{z(1+z)}{1+z^2}.
\end{equation}
Plugging $w(z)$ into the last equality in Eq. (\ref{q}), the parametrization for the deceleration parameter $q$ becomes
\begin{equation}
q(z)=q_0+q_1\frac{z(1+z)}{1+z^2},
\end{equation}
with $q_0=\frac{1}{2}(1+3w_0)$ and $q_1=\frac{3}{2}w_1$. They performed statistical analysis using observational data from Supernova Type Ia (SNLS), Baryon Acoustic Oscillations (SDSS), Cosmic Microwave Background shift parameter (WMAP), and estimates of $H(z)$ (from the ages of high-$z$ galaxies), and obtained the best fit values of the parameters to be $w_0=-1.11$ and $w_1=0.43$ at $1\sigma$ confidence level.

\section{Thermodynamic Consequences}

Our first task is to study the thermodynamics, particularly the GSL and the TE for the above system considering cosmological (or dynamical) apparent horizon as the boundary, since we are interested in the thermodynamic behavior of the local Universe. This type of thermodynamic study was first introduced in their pioneering work by Wang, Gong, and Abdalla \cite{Wang1}. It is worthwhile to mention that while GSL\footnote{The idea of incorporating the GSL in cosmology was first developed by Ram Brustein \cite{Brustein1}. This second law is based on the conjecture that causal boundaries and not only event horizons have geometric entropies proportional to their area.} should hold throughout the evolution of the Universe, TE is required to be satisfied at least during the final phases of its evolution. Now, if $S$ be the entropy of our system (horizon$+$cosmic fluid inside), then GSL and TE, respectively, refer to the inequalities $\dot{S} \geq 0$ and $\ddot{S}<0$. The total entropy will be $S=S_A+S_f$, where $S_A$ and $S_f$ are the entropies of the apparent horizon and the fluid bounded by it respectively. The above two inequalities then get modified as $\frac{d}{dt}(S_A+S_f)\geq 0$ and $\frac{d^2}{dt^2}(S_A+S_f)<0$. As the horizon entropy scales with its surface area, we have
\begin{equation} \label{sa}
S_A=\left(\frac{c^3}{G\hbar}\right) \pi R_{A}^{2},
\end{equation}
where $R_A=\frac{1}{H}$ is the location of the cosmological apparent horizon. The temperature of the horizon is found to be proportional to its surface gravity and goes by the expression
\begin{equation} \label{ta}
T_A=\left(\frac{\hbar c}{\kappa _B}\right) \frac{1}{2\pi R_A}.
\end{equation}
The entropy of the fluid can be obtained from the Gibbs equation
\begin{equation} \label{geq}
T_fdS_f=dE_f+pdV_A,
\end{equation}
where $E_f=\rho V_A$ is the total entropy of the fluid inside the apparent horizon, $V_A=\frac{4}{3}\pi R_{A}^{3}$ is the volume of the fluid, and $\rho$ and $p$ are, respectively, the energy density and the pressure of the cosmic fluid respectively. The fluid temperature is assumed to be equal to the horizon temperature in these kinds of thermodynamic studies, i.e., $T_f=T_A$.

Let us set $8\pi$ and the fundamental constants to unity, without any loss of generality. The the first order time-derivative of the total entropy $S$ can be evaluated as \cite{Wang1,Saha0,Saha00}
\begin{eqnarray} \label{ds}
\dot{S} &=& \frac{d}{dt}(S_A+S_f) \nonumber \\
&=& \frac{9\sqrt{3}}{16\sqrt{\rho}} \left(1+\frac{p}{\rho}\right)^2,
\end{eqnarray}
which shows that the GSL always holds irrespective of the nature of cosmic fluid chosen. Thus, GSL is satisfied throughout the evolution of the Universe.

\begin{figure}[h]
\begin{center}
\begin{minipage}{0.45\textwidth}
\includegraphics[width=1.0\linewidth]{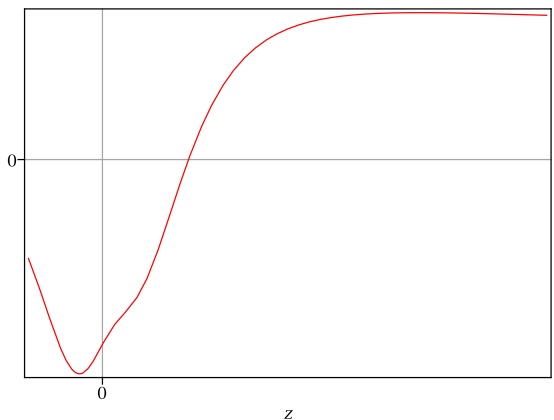}
\end{minipage}
\caption{The variation of $\ddot{S}$ against $z$. We have used the best fit values of $q_0$ and $q_1$ as obtained in Ref. \cite{Barboza1}.}
\label{Fig0}
\end{center}
\end{figure}

The second derivative of $S$ was obtained in the following form \cite{Saha0,Saha00,Saha000}---
\begin{eqnarray} \label{dds}
\ddot{S} &=& \frac{d^2}{dt^2}(S_A+S_f) \nonumber \\
&=& \frac{9}{16}\left(1+\frac{p}{\rho}\right)\left[\left(1+\frac{p}{\rho}\right)\left(1+6\frac{p}{\rho}\right)-\left(5+3\frac{p}{\rho}\right)\frac{\dot{p}}{\dot{\rho}}\right],
\end{eqnarray}
which is not useful for our present investigation. Noting that
$$\frac{\dot{p}}{\dot{\rho}}=\frac{1}{3(1+q)}\left[(1+z)\frac{dq}{dz}+2(1+q)^2-3(1+q)\right],$$
it is remarkable to see that Eq. (\ref{dds}) can be entirely expressed in terms of $q$ and $\frac{dq}{dz}$ as
\begin{equation}
\ddot{S} = \frac{1}{4}\left[(1+q)(1+2q^2)-(2+q)(1+z)\frac{dq}{dz}\right]
\end{equation}
using the Einstein field equations and the definition of the deceleration parameter, i.e., the first equality of Eq. (\ref{q}). In order to know how $\ddot{S}$ behaves during the last stages of the evolution, we have plotted it against $z$ with the help of Maple plotting software (see Figure \ref{Fig0}). It is clearly visible from the figure that our physical system remains in TE during the final phases of evolution of the Universe. Thus, the Barboza-Alcaniz EoS paramatrization and the resulting $q$-paramterization is well consistent with GSL and TE and therefore can be considered to be thermodynamically motivated. Moreover, our deductions are perfectly in line with previous studies undertaken in this direction \cite{Radicella1,Mimoso1,Pavon1}.

\section{Implications of isentropic particle creation mechanism}

We now introduce a dissipative effect (in the FLRW Universe) in the form of a bulk viscous pressure $\Pi$ which occurs in the expression for the energy-momentum tensor in the following manner ---
\begin{equation}
T_{\mu \nu}=(\rho +p+\Pi)u_{\mu}u_{\nu}+(p+\Pi)g_{\mu \nu}.
\end{equation}
In Cosmology, it is customary to assume that the dissipation arises solely due to the nonconservation of the (quantum) particle number \cite{Zeldovich1,Murphy1,Hu1}, in other words, due to isentropic (or adiabatic) production of perfect fluid particles \cite{Prigogine1,Calvao1}. The associated Friedmann and acceleration equations are
\begin{equation} \label{fa1}
H^2=\frac{8\pi G}{3}\rho~~~~~~~~\text{and}~~~~~~~~\dot{H}=-4\pi G(\rho +p+\Pi).
\end{equation}
The conservation laws, namely, 
\begin{equation}
T_{;\nu}^{\mu \nu}=0~~~~\text{and}~~~~N_{;\mu}^{\mu}=0
\end{equation}
have the explicit expressions
\begin{equation} \label{emce1}
\dot{\rho}+\theta(\rho +p+\Pi)=0~~~~~~~~\text{and}~~~~~~~~\dot{n}+\theta n=0.
\end{equation}
Here, $\theta =u_{;\mu}^{\mu}$ is the fluid expansion, $N^{\mu}=nu^{\mu}$ is the particle flow vector, $n$ is the particle number density, and $\dot{n}=n_{,\alpha}u^{\alpha}$.

The phenomenon of nonconservation of the total number $N$ of particles in an open thermodynamic system modifies the second expression in Eq. (\ref{emce1}) as
\begin{equation} \label{ngamma}
\dot{n}+\theta n=n\Gamma,
\end{equation}
where $\Gamma$ denotes the rate of change of the number of particles ($N=na^3$) in a comoving volume $a^3$. Hence, $\Gamma >0$ refers to creation of particles while $\Gamma <0$ indicates particle annihilation. In this scenario, the energy density $\rho$ and the thermostatic pressure $p$ of the cosmic fluid is assumed to be related by the EoS $p=(\gamma -1)\rho$ with $\frac{2}{3}\leq \gamma \leq 2$.  The bounds on $\gamma$ ensures that the fluid EoS does not become exotic.

Now, using the first expression in Eq. (\ref{emce1}), Eq. (\ref{ngamma}), and the Gibbs relation ($T$ is the temperature and $s$ is the entropy per particle)
\begin{equation}
Tds=d\left(\frac{\rho}{n}\right)+pd\left(\frac{1}{n}\right),
\end{equation}
we obtain
\begin{equation}
\Pi=-\frac{\Gamma}{\theta}(\rho +p)
\end{equation}
under the assumption of isentropicity\footnote{Note that although the entropy per particle is constant, there is entropy production due to the enlargement of the phase space of the system.}, i.e., $\dot{s}=0$. Plugging the above expression for $\Pi$ in the Einstein field equation (\ref{fa1}), we get
\begin{eqnarray}
\frac{\Gamma}{\theta}=1+\left(\frac{2}{3\gamma}\right)\frac{\dot{H}}{H^2}.
\end{eqnarray}
It is evident from the first equality in Eq. (\ref{q}) that the particle creation rate $\Gamma$ can be written in terms of the deceleration parameter $q$ as \cite{Saha1}
\begin{equation}
\Gamma =3H\left[1-\frac{2}{3\gamma}(1+q)\right],
\end{equation}
where the Hubble parameter $H$ is given by
\begin{equation}
H=H_0\text{exp}\left[\int _{0}^{z}\left(\frac{1+q}{1+z}\right)dz \right],
\end{equation}
where $H_0$ is the value of the Hubble parameter at the present epoch. Note that $\theta =3H$ in a FLRW universe. %It was shown in Ref. \cite{Saha1} that the entropy $S$ and the temperature $T$ in the gravitationally induced particle creation scenario can be expressed in terms of q as
%\begin{equation}
%S=S_0\text{exp}\left[3\int_{z}^{0}\left\{1-\frac{2}{3\gamma}(1+q)\right\}\frac{dz}{1+z}\right]
%\end{equation}
%and
%\begin{equation}
%T=T_1\left[S(1+z)^3\right]^{\gamma -1}, ~~~~~~\text{with}~~T_1=\frac{T_0}{S_{0}^{\gamma -1}}
%\end{equation}
%respectively with $S_0$ and $T_0$ as the values at $z=0$.

Now, taking the form of $q$ obtained with the Barboza-Alcaniz EoS parametrization, we have obtained the expressions for $H$ and $\Gamma$ in terms of the redshift $z$ as
\begin{eqnarray}
H(z) &=& H_0 (1+z)^{1+q_0}(1+z^2)^{\frac{q_1}{2}}, \\ \nonumber \\
\Gamma (z) &=& 3H_0 (1+z)^{1+q_0}(1+z^2)^{\frac{q_1}{2}} \left[1-\frac{2}{3\gamma}\left\{(1+q_0)+q_1\frac{z(1+z)}{1+z^2}\right\}\right].%, \\ \nonumber \\
%S(z) &=& S_0 \left[\frac{(1+z^2)^{\frac{q_1}{3\gamma}}}{(1+z)^{1-\frac{2}{3\gamma}(1+q_0)}}\right]^3, \\ \nonumber \\
%T(z) &=& T_0 \left[(1+z^2)^{\frac{q_1}{3\gamma}}(1+z)^{\frac{2}{3\gamma}(1+q_0)}\right]^{\gamma -1}.
\end{eqnarray}

\begin{figure}[h]
\begin{center}
\begin{minipage}{0.45\textwidth}
\includegraphics[width=0.9225\linewidth]{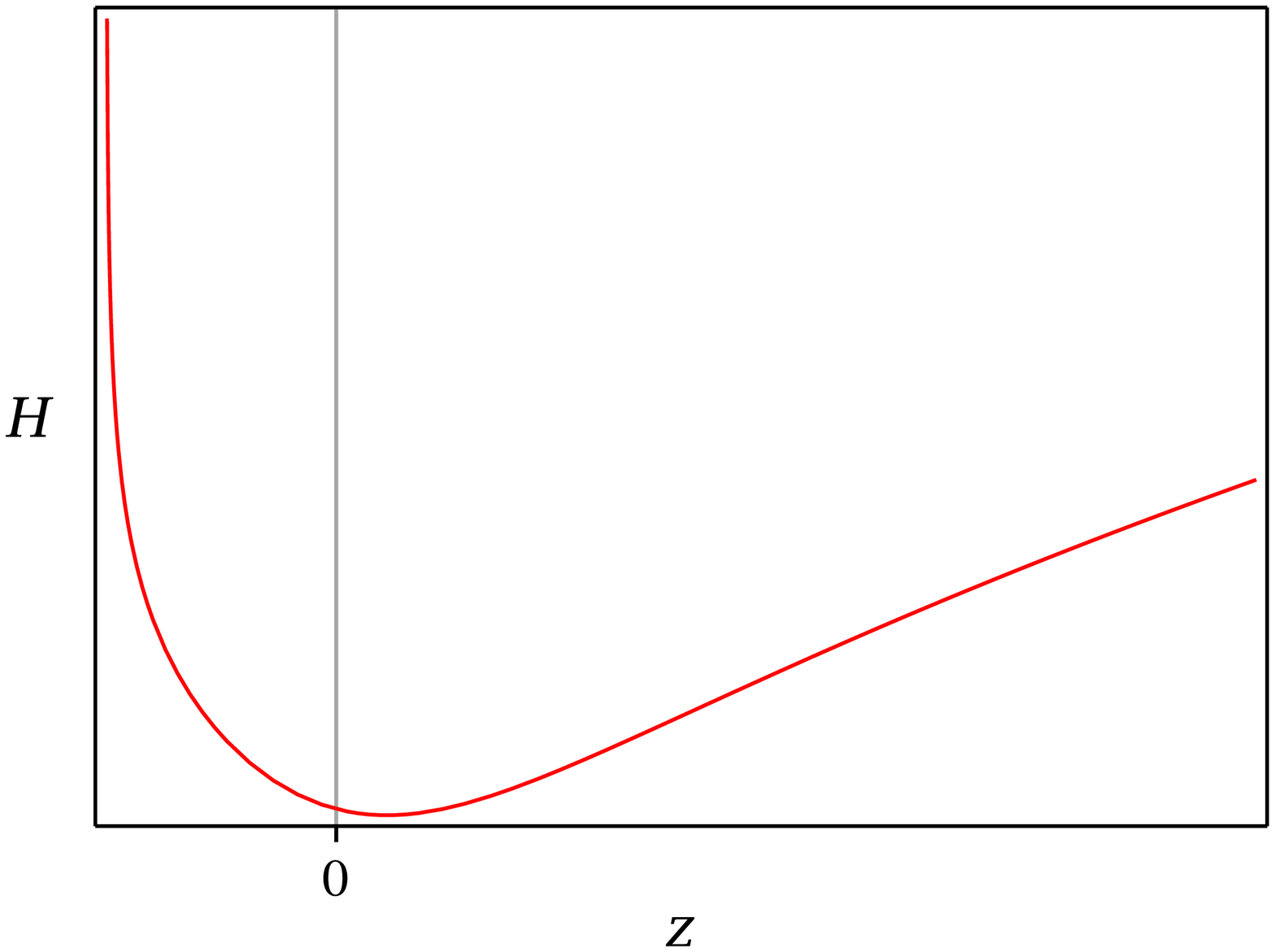}
\end{minipage}
\hspace*{0.6cm}
\begin{minipage}{0.45\textwidth}
\includegraphics[width=0.9225\linewidth]{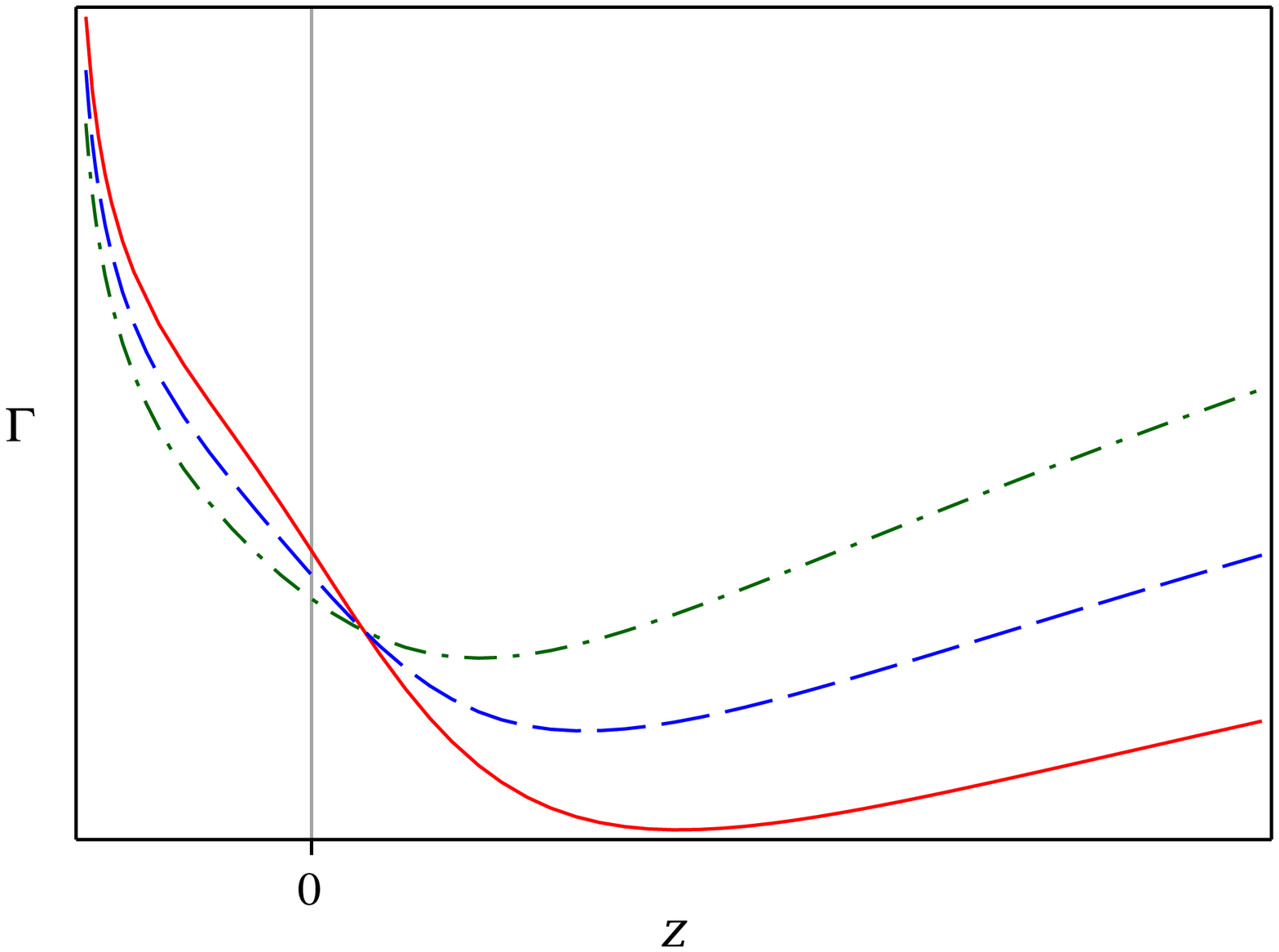}
\end{minipage}
\caption{\textsc{Left Panel:} Variation of $H$ against $z$ and \textsc{Right Panel:} Variation of $\Gamma$ against $z$ for $\gamma =\frac{2}{3}$ (red solid curve), $\gamma =1$ (blue dashed curve), and $\gamma =2$ (green dashdot curve). We have used the best fit values of $q_0$ and $q_1$ as obtained in Ref. \cite{Barboza1}. $H_0$ has been set to $1$.}
\label{Fig1}
\end{center}
\end{figure}

The behavior of the particle creation rate $\Gamma$ is similar for all the three choices of the perfect fluid EoS parameter $\gamma$. Now, thermodynamic arguments tell us that \cite{Gunzig1} in the very early Universe, starting from a regular vacuum, most of the particle creation effectively took place and the creation rate was greater than the expansion rate, while particle creation was strongly suppressed during the radiation era \cite{Parker1}. Also, the creation rate again dominates during the late time acceleration phase. The right panel of Figure \ref{Fig1} exactly demonstrates the above situations during the various phases of evolution of the Universe. For $z \gg 1$ the creation rate is greater and it boils down to a very small value during the deceleration phase. The recent acceleration phase again sees a dominance of the creation rate. Further, it predicts a significant increase in the creation of particles during future phases of evolution. Thus, the Barboza-Alcaniz EoS parametrization is consistent with the adiabatic particle creation mechanism.

\section{Short discussion}

The paper dealt with an investigation of the Barboza-Alcaniz EoS parametrization in terms of thermodynamic motivation as well as correspondence with the gravitationally induced adiabatic particle creation scenario. It has been found that the associated $q$-parametrization is well consistent with the GSL (throughout the evolution) as well as TE (during the final phases of evolution). Also, our study is perfectly in line with previous investigations in this direction. From the expression connecing the deceleration parameter and the particle creation rate $\Gamma$ in an adiabatic particle creation mechanism, we have expressed $\Gamma$ in terms of the deceleration parameter $z$. We have plotted the variation of $\Gamma$ for three values of the perfect fluid EoS parameter $\gamma$ using Maple and found that it's behavior is consistent during each phase of evolution of the Universe as speculated from thermodynamic arguments. Our analysis has also predicted a significant increase in the creation of particles during the final phases of evolution of the Universe.

%%%%%%%%%%%%%%%%%%%%%%%%%%%%%%%%%%%%%%%%%%%%%%%%%%%%%%%%%%%%%%%%%%%%%%%%%%%%%%%%%%%%%%%%%%%%%%%%%%%%%%%%%%%%%%%%%%%%%%%%%%%%

\begin{acknowledgments}
AM wishes to thank DST, Govt. of India for providing a research fellowship. SS was partially supported by SERB, Govt. of India under National Post-doctoral Fellowship Scheme [File No. PDF/2015/000906].
\end{acknowledgments}

%%%%%%%%%%%%%%%%%%%%%%%%%%%%%%%%%%%%%%%%%%%%%%%%%%%%%%%%%%%%%%%%%%%%%%%%%%%%%%%%%%%%%%%%%%%%%%%%%%%%%%%%%%%%%%%%%%%%%%%%%%%%


\frenchspacing
\begin{thebibliography}{150}

\bibitem{q1} M. S. Turner and A. G. Riess, {\color{magenta} Astrophys. J., {\bf 569}, 18 (2002)}. 
\bibitem{q2} A. G. Riess et al., {\color{magenta} Astrophys. J., {\bf 607}, 665 (2004)}.
\bibitem{q3} Y. G. Gong and A. Wang, {\color{magenta} Phys. Rev. D, {\bf 73}, 083506 (2006)}.
\bibitem{q0} Y. Gong and A. Wang, {\color{magenta} Phys. Rev. D, {\bf 75}, 043520 (2007)}.
\bibitem{q4} J. V. Cunha and J. A. S. Lima, {\color{magenta} Mon. Not. R. Astron. Soc., {\bf 390}, 210 (2008)}.
\bibitem{q5} J. V. Cunha, {\color{magenta} Phys. Rev. D, {\bf 79}, 047301 (2009)}.
\bibitem{q6} B. Santos, J. C. Carvalho, and J. S. Alcaniz, {\color{magenta} Astropart. Phys. {\bf 35}, 17 (2011)}. 
\bibitem{q7} R. Nair et al. {\color{magenta} JCAP, {\bf 01}, 018 (2012)}.
\bibitem{q8} O. Akarsu et al., {\color{magenta} Eur. Phys. J. Plus, {\bf 129}, 22 (2014)}.
\bibitem{q9} L. Xu and H. Liu, {\color{magenta} Mod. Phys. Lett. A, {\bf 23}, 1939 (2008)}.
\bibitem{q10} L. Xu and J. Lu, {\color{magenta} Mod. Phys. Lett. A, {\bf 24}, 369 (2009)}.
\bibitem{q11} S. del Campo et al., {\color{magenta} Phys. Rev. D, {\bf 86}, 083509 (2012)}.
\bibitem{q12} A. A. Mamon and S. Das, {\color{magenta} Int. J. Mod. Phys. D, {\bf 25}, 1650032 (2016)}.
\bibitem{q13} A. A. Mamon and S. Das, arXiv: 1610.07337 [gr-qc].
\bibitem{Barboza1} E. M. Barboza Jr. and J. S. Alcaniz, {\color{magenta} Phys. Lett. B {\bf 666}, 415 (2008)}.
\bibitem{Wang1} B. Wang, Y. Gong, and E. Abdalla, {\color{magenta} Phys. Rev. D {\bf 74}, 083520 (2006)}.
\bibitem{Brustein1} R. Brustein, {\color{magenta} Phys. Rev. Lett. {\bf 84}, 2072 (2000)}.
\bibitem{Saha0} S. Saha and S. Chakraborty, {\color{magenta} Phys. Lett. B {\bf 717}, 319 (2012)}.
\bibitem{Saha00} S. Saha and S. Chakraborty, {\color{magenta} Phys. Rev. D {\bf 89}, 043512 (2014)}.
\bibitem{Saha000} S. Saha and A. Mondal, {\color{magenta} Eur. Phys. J. C {\bf 77}, 196 (2017)}.
\bibitem{Radicella1} N. Radicella and D. Pavon, {\color{magenta} Gen. Relativ. Gravit. {\bf 44}, 685 (2012)}.
\bibitem{Mimoso1} J.P. Mimoso and D. Pavon, {\color{magenta} Phys. Rev. D {\bf 87}, 047302 (2013)}.
\bibitem{Pavon1} D. Pavon, {\color{magenta} Int. J. Geom. Methods Mod. Phys. {\bf 11}, 1460007 (2014)}.
\bibitem{Zeldovich1} Ya B. Zel'dovich, {\color{magenta} JETP Lett. {\bf 12}, 307 (1970)}.
\bibitem{Murphy1} G. L. Murphy, {\color{magenta} Phys. Rev. D {\bf 8}, 4231 (1973)}.
\bibitem{Hu1} B. L. Hu, {\color{magenta} Phys. Lett. A {\bf 90}, 375 (1982)}.
\bibitem{Prigogine1} I. Prigogine, J. Geheniau, E. Gunzig, and P. Nardone, {\color{magenta} Gen. Relativ. and Gravit. {\bf 21}, 767 (1989)}.
\bibitem{Calvao1} M. O. Calvao, J. A. S. Lima, and I. Waga, {\color{magenta} Phys. Lett. A {\bf 162}, 223 (1992)}.
\bibitem{Saha1} S. Saha and S. Chakraborty, {\color{magenta} Gen. Relativ. Gravit. {\bf 47}, 127 (2015)}.
\bibitem{Gunzig1} E. Gunzig, R. Maartens, and A. V. Nesteruk, {\color{magenta} Class. Quantum Grav. {\bf 15}, 923 (1998)}.
\bibitem{Parker1} L. Parker and D. J. Toms, {\it Quantum Field Theory in Curved Spacetime: Quantized Field and Gravity} (Cambridge Univ. Press, Cambridge, England, 2009).

%\bibitem{Birrell1} N.D. Birrell and P.C.W. Davies, {\it Quantum Fields in Curved Space} (Cambridge Univ. Press, Cambridge, England, 1982).

%\bibitem{Lima00} J.A.S. Lima, M.O. Calvao, and I. Waga, arXiv: 0708.3397.


\end{thebibliography}
\end{document}